\documentclass[journal=jacsat]{achemso}

\usepackage[T1]{fontenc}

\usepackage{amsmath,color}
\usepackage{amssymb}
\usepackage{amsfonts}
\usepackage{mathtools}
\usepackage{physics}

\usepackage[labelfont=bf]{caption}
\usepackage[labelsep=period]{caption}
\usepackage{threeparttable}
\usepackage{tabularx}
\usepackage{titlesec}
\usepackage{ragged2e}
\usepackage{placeins}

\usepackage{orcidlink}
\hypersetup{hidelinks}

\newcommand{\mb}[1]{\mathbf{#1}}
\newcommand{\mc}[1]{\mathcal{#1}}
\newcommand{\ur}{\ensuremath{_{\text{R}}}}
\newcommand{\up}{\ensuremath{_{\text{P}}}}
\newcommand{\uda}{\ensuremath{_{\text{DA}}}}
\newcommand{\uc}{\ensuremath{_{\text{c}}}}
\newcommand{\dt}{\ensuremath{\Delta t}}
\newcommand{\gp}{\ensuremath{\mathbf{g}_{\perp}}}
\newcommand{\gv}{\nabla_{\mb{R}}V(\mb{R}_{\text{MECP}})}
\newcommand{\ud}{\ensuremath{_{\text{D}}}}
\newcommand{\ua}{\ensuremath{_{\text{A}}}}

\title{\large Diabatic Seam Space Sampling for Hydrogen Tunneling \\ Systems with Nuclear--Electronic Orbital Theory}

\author{\normalsize Joseph A. Dickinson\orcidlink{0000-0002-5601-7050}}
\affiliation{\footnotesize Department of Chemistry, Yale University, New Haven, CT 06520, USA}
\alsoaffiliation{\footnotesize Department of Chemistry, Princeton University, Princeton, NJ 08544, USA}

\author{\normalsize Eno Paenurk\orcidlink{0000-0002-6921-757X}}
\affiliation{\footnotesize Fakultät für Chemie und Pharmazie, Universität Regensburg, 93053 Regensburg, Germany}
\alsoaffiliation{\footnotesize Department of Chemistry, Princeton University, Princeton, NJ 08544, USA}

\author{\normalsize Sharon Hammes-Schiffer\orcidlink{0000-0002-3782-6995}}
\email{shs566@princeton.edu}
\affiliation{\footnotesize Department of Chemistry, Princeton University, Princeton, NJ 08544, USA}
	
\begin{document}

\makeatletter
\renewcommand{\eqref}[1]{\ref{#1}}
\makeatother

\begin{abstract}
\begin{spacing}{1.0}
\noindent 
Hydrogen tunneling is central to many chemical and biological processes. 
Herein, we introduce the vibronic-SHAKE (V-SHAKE) approach to comprehensively sample energy-conserving molecular configurations enabling hydrogen tunneling.
A constrained form of nuclear--electronic orbital multistate density functional theory (NEO-MSDFT) dynamics, where the tunneling hydrogen nucleus is quantized, is used to sample geometries in the diabatic seam space corresponding to the intersection of the reactant and product NEO-DFT diabatic vibronic surfaces.
V-SHAKE is applied to hydrogen and deuterium tunneling in 4-cyanobutanolate and (\textit{Z})-4-hydroxybut-3-en-2-one.
The vibronic coupling is found to vary significantly in the diabatic seam space, mainly due to changes in the donor--acceptor distance. 
The reaction coordinate and gradient of the vibronic coupling at the minimum energy crossing point are nearly orthogonal and are dominated by motions stabilizing the product relative to the reactant or decreasing the donor--acceptor distance, respectively.
V-SHAKE provides fundamental insights and validation for assumptions underlying rate theories.

\vspace{2.0em} 

\noindent \textbf{\normalsize TOC Graphic}
\bigskip
\FloatBarrier
\includegraphics[width=3.0in, height=1.65in]{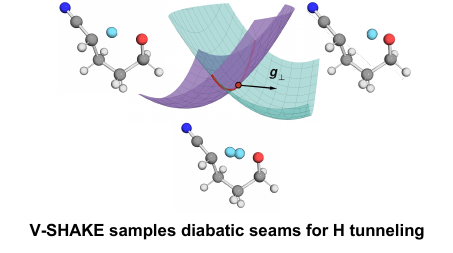} 
\end{spacing}
\end{abstract}
\maketitle 
\pagebreak

Hydrogen tunneling underpins a large number of chemical and biological processes. \cite{cha_hydrogen_1989, hammes-schiffer_hydrogen_2006, richardson_concerted_2016, vaillant_tunneling_2019, litman_elucidating_2019} 
The measurement of kinetic isotope effects (KIEs) can be used to probe reaction mechanisms \cite{anslyn_modern_2006, gomez-gallego_kinetic_2011, simmons_interpretation_2012, truong_large_2021} across a wide array of chemical processes \cite{nagel_tunneling_2006, edwards_analysis_2009, klinman_understanding_2018, hammes-schiffer_explaining_2025} and are often used to analyze the degree to which hydrogen tunneling affects the observed rate constants.
This significance has motivated the development of methods to describe hydrogen tunneling systems and compute KIEs, such as wavepacket \cite{meyer_multi-configurational_1990, beck_multiconfiguration_2000, meyer_quantum_2003, hammer_multiconfigurational_2009} and path integral approaches. \cite{berne_classical_1998, tuckerman_path_2002, craig_quantum_2004, habershon_ring-polymer_2013}
Despite the accuracy of these methods, applications to sizable molecular systems are challenging, motivating the development of methodologies in which hydrogen tunneling and KIEs can be investigated in a computationally efficient manner. 

The nuclear--electronic orbital (NEO) method \cite{webb_multiconfigurational_2002, pavosevic_multicomponent_2020, hammes-schiffer_nuclearelectronic_2021} is a multicomponent quantum chemistry approach in which select nuclei, typically protons or deuterons, are quantized and treated at the same level as the electrons. 
Many methods have been developed within the NEO framework, \cite{pavosevic_multicomponent_2019, hammes-schiffer_nuclearelectronic_2021, alaal_multicomponent_2021, hasecke_local_2024, malbon_nuclearelectronic_2025, stein_computing_2025, goudy_triple_2025} with NEO density functional theory (NEO-DFT) \cite{pak_density_2007, brorsen_multicomponent_2017, yang_development_2017, yang_multicomponent_2018} emerging as the main NEO approach for large-scale chemical applications \cite{chow_nuclear-electronic_2023, li_nuclear-electronic_2023, chow_nuclear_2024, smith_isotope_2025, smith_capturing_2026} and proton transfer dynamics simulations. \cite{tao_direct_2021, zhao_real-time_2020, zhao_excited_2021, li_semiclassical_2022, chow_nuclearelectronic_2023, dickinson_extended_2026}
NEO multistate DFT (NEO-MSDFT) \cite{yu_nuclear-electronic_2020, yu_analytical_2022, dickinson_generalized_2023} was introduced to describe hydrogen tunneling systems within the NEO-DFT framework.
The NEO-MSDFT approach has been shown to produce accurate tunneling splittings and proton densities at fixed geometries for a wide array of single-proton\cite{yu_nuclear-electronic_2020} and multi-proton \cite{dickinson_generalized_2023} systems. 
It has also been used to simulate nonadiabatic hydrogen tunneling dynamics in malonaldehyde \cite{yu_nonadiabatic_2022} and doubly H-bonded dimers. \cite{dickinson_nonadiabatic_2024} 
Thus, the NEO-MSDFT method provides a natural and efficient framework for describing hydrogen tunneling and investigating the associated KIEs. 

Recently, a NEO general rate theory (NEO-GRT) was introduced to compute primary H/D KIEs spanning vibrationally adiabatic and nonadiabatic regimes for electronically adiabatic deep tunneling systems. \cite{paenurk_nuclearelectronic_2026}
NEO-GRT was shown to predict KIEs that are in agreement with ring-polymer instanton theory \cite{richardson_ring-polymer_2009, richardson_perspective_2018, richardson_ring-polymer_2018} at temperatures as low as 50~K for full-dimensional molecular systems. 
In this theory, the rate constant for a hydrogen transfer reaction is formulated in terms of reactant and product diabatic vibronic states, corresponding to the transferring proton or deuteron density localized near its donor or acceptor, respectively. 
For many systems, particularly in the vibrationally nonadiabatic regime, the rate constant is strongly influenced by the vibronic coupling between the reactant and product diabatic vibronic states. 
Because the vibronic coupling between the reactant and product diabatic vibronic states varies strongly with the donor--acceptor distance, the overall rate constant is computed by calculating the rate constant at a series of donor--acceptor distances and thermally averaging over this distance.

\begin{figure}
    \centering 
    \includegraphics[width=4.5in]{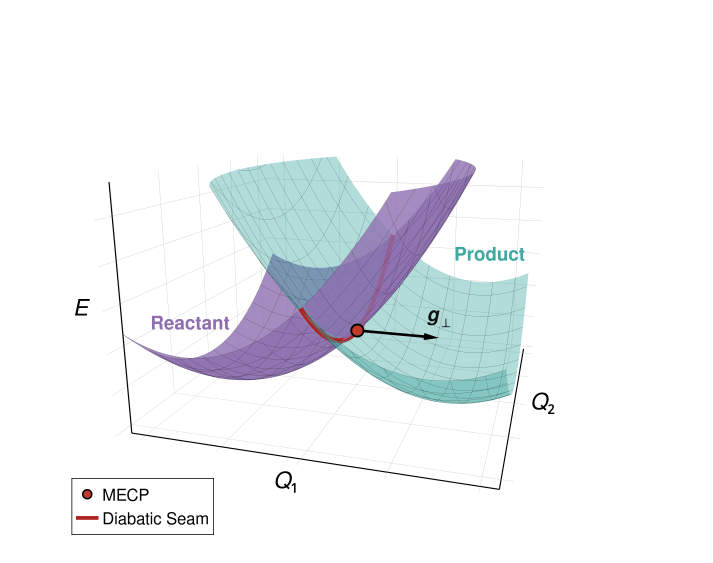}
    \caption{
    Schematic of two intersecting reactant (purple) and product (teal) diabatic vibronic surfaces and the associated diabatic seam. 
    The energy, $E$, of each state is plotted as a function of two arbitrary coordinates, $Q_1$ and $Q_2$. 
    Since the diabatic vibronic surfaces are two-dimensional surfaces, their diabatic seam (shown in red) is a one-dimensional line. 
    The minimum energy crossing point (MECP), defined as the point of minimum energy on the diabatic seam, is labeled by a red dot. 
    The vector $\gp$, defined as the difference in the diabatic energy gradients at the MECP, is perpendicular to the diabatic seam and corresponds to the reaction coordinate pointing from the reactant to the product state. 
    In this work, the NEO-DFT reactant and product diabatic states are defined by the localization of the proton/deuteron density closer to the donor or acceptor atom, respectively. 
    The diabatic seam between two NEO-DFT diabatic vibronic surfaces contains all tunneling-ready nuclear configurations in hydrogen tunneling systems. 
    } 
    \label{fig:seam_schematic}
\end{figure}

Within theories such as NEO-GRT, the vibronic coupling is evaluated at molecular structures for which these states are equal in energy.
These so-called "tunneling-ready" \cite{roston_critical_2013, roston_kinetic_2014, sakhaee_hydride_2019, bai_rigidity_2025} structures exist in the diabatic seam space, \cite{cofer-shabica_marcus_2026} defined as the submanifold of configuration space in which the diabatic vibronic states are isoenergetic to enable hydrogen tunneling \cite{paenurk_nuclearelectronic_2026} (see Figure~\ref{fig:seam_schematic}).
This seam dictates NEO-GRT input quantities and contains all tunneling-ready nuclear configurations. 
Thus, the development of methods to characterize these seams is desirable.
Such methods also enable an analysis of the validity of the Condon approximation\cite{medvedev_inelastic_1997, jang_theory_2005} in these types of systems, clarifying whether the vibronic coupling remains predominantly constant in the relevant crossing region (Condon behavior) or varies significantly in this region (non-Condon behavior). 
Moreover, if the Condon approximation is violated, analysis of the dominant coordinate(s) leading to this variation in the vibronic coupling identifies coordinates that should be treated explicitly in theories such as NEO-GRT.

\begin{figure}
    \centering 
    \includegraphics[width=6.5in]{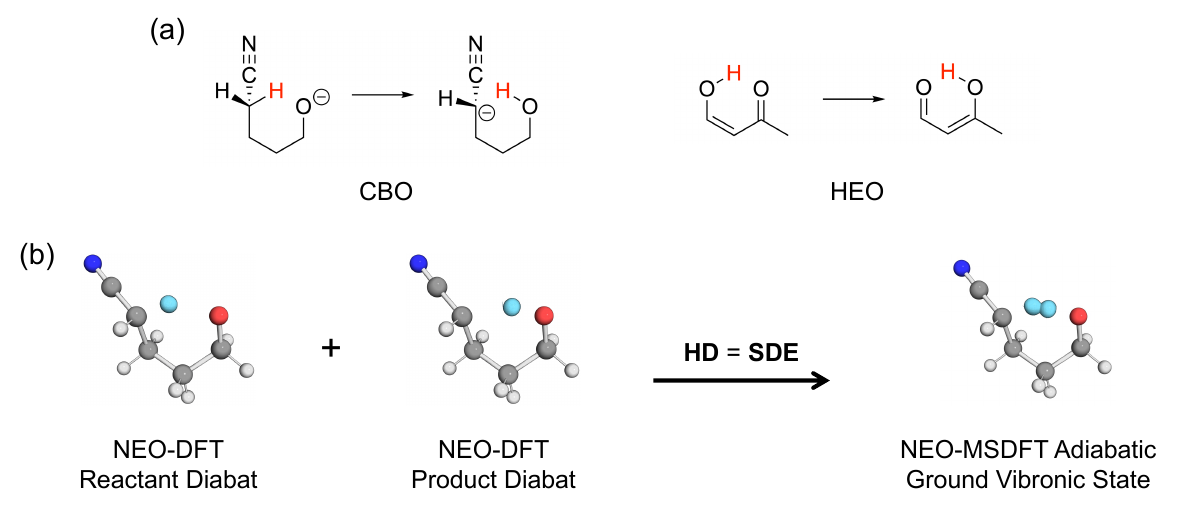}
    \caption{
    (a) The two hydrogen tunneling systems studied in this work: 4-cyanobutanolate (CBO) and (\textit{Z})-4-hydroxybut-3-en-2-one (HEO). 
    The tunneling hydrogen that was isotopically substituted is shown in red and is the only quantized nucleus in each system. 
    (b) Schematic of the NEO-MSDFT method for CBO.
    The reactant and product NEO-DFT diabatic vibronic states, defined by the proton or deuteron density localized near the donor or acceptor atom, respectively, are linearly combined to obtain the NEO-MSDFT adiabatic vibronic states.
    The proton densities are shown in cyan.
    } 
    \label{fig:lewis_dens}
\end{figure}

Herein, we introduce a methodology to comprehensively explore and characterize these diabatic seam spaces. 
This work is inspired by the electronic-SHAKE (E-SHAKE) approach,\cite{cofer-shabica_marcus_2026} as well as other methods that sample or optimize on seam spaces for electronic conical intersections.\cite{levine_optimizing_2007, mori_exploring_2013, lindner_metafalcon_2019} 
We formulate a constrained adiabatic dynamics approach on the NEO-MSDFT ground vibronic state that can be used to sample the diabatic seam space between the NEO-DFT reactant and product vibronic surfaces. 
As the diabatic seam space between NEO-DFT vibronic states is sampled, we name this approach vibronic-SHAKE (V-SHAKE). 
After presenting the theory underlying V-SHAKE, we use this approach to sample the diabatic seam spaces of two compounds for both H and D isotopes: 4-cyanobutanolate (CBO), one of the deep tunneling systems studied in ref~\citenum{paenurk_nuclearelectronic_2026}, as well as (\textit{Z})-4-hydroxybut-3-en-2-one (HEO), a methyl-substituted form of malonaldehyde (Figure~\ref{fig:lewis_dens}a). 

We begin with an overview of the NEO-MSDFT approach for the case of a single tunneling hydrogen nucleus \cite{yu_nuclear-electronic_2020}.
Only this case is studied herein, although the NEO-MSDFT approach has been generalized to systems with multiple tunneling hydrogen nuclei. \cite{dickinson_generalized_2023}
In NEO-MSDFT, two localized diabatic NEO-DFT vibronic states are defined: the reactant state, $\ket*{\tilde{\Psi}\ur}$, and the product state, $\ket*{\tilde{\Psi}\up}$. 
The reactant and product diabatic vibronic states are defined by the proton or deuteron density localizing closer to the donor or acceptor atom, respectively (Figure~\ref{fig:lewis_dens}b). 
The adiabatic NEO-MSDFT vibronic states are expanded in terms of these localized diabatic NEO-DFT vibronic states in a nonorthogonal configuration interaction (NOCI) scheme \cite{skone_nuclear-electronic_2005, thom_hartreefock_2009} according to
\begin{equation}\label{eq:msdft_expansion}
\begin{aligned}
    \ket{\Psi_0} &= D\ur^0 \ket*{\tilde{\Psi}\ur} + D\up^0 \ket*{\tilde{\Psi}\up} \\
    \ket{\Psi_1} &= D\ur^1 \ket*{\tilde{\Psi}\ur} + D\up^1 \ket*{\tilde{\Psi}\up} \\
\end{aligned}
\end{equation}
Here, $\ket{\Psi_0}$ and $\ket{\Psi_1}$ are the adiabatic NEO-MSDFT ground and excited vibronic states, respectively, and $D_{\text{R(P)}}^{i}$ is the coefficient of the reactant (product) diabatic state for the $i$-th NEO-MSDFT state.

The $D_{\text{R(P)}}^{i}$ coefficients in eq~\eqref{eq:msdft_expansion} are found by solving the $2\times 2$ generalized eigenvalue problem
\begin{equation}\label{eq:msdft_eigenproblem}
    \mathbf{H}\mathbf{D} = \mathbf{S}\mathbf{D}\mathbf{E}
\end{equation}
where $\mathbf{H}$ and $\mathbf{S}$ are the NEO-MSDFT Hamiltonian and overlap matrices, respectively, $\mathbf{D}$ is the matrix of eigenvectors containing the coefficients $D_{\text{R(P)}}^{i}$, and $\mathbf{E}$ is the diagonal matrix containing the adiabatic NEO-MSDFT ground and excited vibronic state energies, $E_0$ and $E_1$, respectively. 
The diagonal elements of $\mathbf{H}$ are the NEO-DFT energies associated with $\ket*{\tilde{\Psi}\ur}$ and $\ket*{\tilde{\Psi}\up}$, denoted $E\ur$ and $E\up$, respectively. 
The off-diagonal elements of $\mathbf{H}$ are represented by a physically motivated form inspired by electronic MSDFT. \cite{mo_energy_2011, gao_beyond_2016, grofe_diabatic-at-construction_2017, lu_multistate_2022}
For geometries at which $E\ur \approx E\up$ and the proton/deuteron density is bilobal (adiabatic ground vibronic state in Figure~\ref{fig:lewis_dens}b), the tunneling splitting, $\Delta$, is defined as the difference between the adiabatic NEO-MSDFT energies, i.e., $\Delta=E_1-E_0$.  
When $E\ur=E\up$, the vibronic coupling, $V$, is half the tunneling splitting, i.e., $V=\Delta/2$. 

The expressions for the individual matrix elements in eq~\eqref{eq:msdft_eigenproblem} and the analytical gradients are provided elsewhere.\cite{yu_nuclear-electronic_2020, dickinson_generalized_2023,yu_analytical_2022, dickinson_nonadiabatic_2024} 
Additional details about the NEO-MSDFT approach are provided in Section~\ref{sec:si_bl_msdft} of the SI. 
Specifically, this section provides details about the block-localization \cite{cembran_block-localized_2009} procedure for NEO-MSDFT used in this work. 
This section also discusses the two parameters utilized in NEO-MSDFT\cite{yu_nuclear-electronic_2020, dickinson_generalized_2023} as well as the reparameterization performed for this work. 
We note that the NEO-MSDFT method is based on the premise that the transferring hydrogen moves on a double-well Born--Oppenheimer potential energy surface, providing a clear definition of the reactant and product diabatic vibronic states in terms of the proton localized in each of the wells. 
When the transferring hydrogen moves on a single-well or nearly single-well Born--Oppenheimer potential energy surface, the reactant and product diabatic vibronic states are not well-defined.  
Such structures are not on the diabatic seam and do not get sampled by V-SHAKE. 

We now provide the equations-of-motion underlying the V-SHAKE approach, using notation similar to that used in ref~\citenum{cofer-shabica_marcus_2026}. 
Let $\mb{R}$ represent the $3N\uc$-dimensional vector of classical nuclear positions, where $N\uc$ is the number of classical nuclei.  
The classical nuclear dynamics are governed by the V-SHAKE Lagrangian, $\mc{L}_{\text{V-SHAKE}}$, given by
\begin{equation}\label{eq:lagr_vshake}
    \mc{L}_{\text{V-SHAKE}}(\mb{R},\dot{\mb{R}},\gamma) = \frac{1}{2}\dot{\mb{R}}^{\text{T}}\mb{M}\dot{\mb{R}}-E_0(\mb{R})-\gamma\sigma(\mb{R})
\end{equation}
Here, $\mb{M}$ is the $3N\uc\times3N\uc$ diagonal mass matrix, with the mass of each nucleus repeated along the three Cartesian elements associated with that nucleus, and $E_0(\mb{R})$ is the energy of the adiabatic NEO-MSDFT ground vibronic state. Moreover, $\sigma(\mb{R})$ is the difference between the reactant and product diabatic NEO-DFT energies at $\mb{R}$
 \begin{equation}\label{eq:sigma_def}
     \sigma(\mb{R}) = E\ur(\mb{R}) - E\up(\mb{R})
 \end{equation}
and $\gamma$ is a Lagrange multiplier that enforces the constraint $\sigma(\mb{R}) = 0$ throughout the dynamics. 
The objective of V-SHAKE is to sample geometries in the diabatic seam space, where $E\ur=E\up$, thereby requiring the enforcement of the constraint $\sigma(\mb{R})=0$ during V-SHAKE dynamics. 

Application of the Euler-Lagrange equation to $\mc{L}_{\text{V-SHAKE}}$ of eq~\eqref{eq:lagr_vshake} yields the constrained equations-of-motion $\mb{M}\ddot{\mb{R}} = \mb{F}(\mb{R}) + \gamma\mb{G}(\mb{R})$ subject to $\sigma(\mb{R})=0$, where $\mb{F}=-\nabla_{\mb{R}}E_0(\mb{R})$ and $\mb{G}=-\nabla_{\mb{R}}\sigma(\mb{R})$.
We integrate the constrained equations-of-motion using the velocity-Verlet \cite{swope_computer_1982} algorithm to obtain equations analogous to those used in the standard RATTLE\cite{andersen_rattle_1983} algorithm:
\begin{subequations}\label{eq:vshake_int_eom}
\begin{align}
    \mb{R}(t+\dt) &= \mb{R}(t) + \dt\dot{\mb{R}}(t) + \frac{\dt^2}{2}\mb{M}^{-1}\left[\mb{F}(t) + \gamma_{x}\mb{G}(t)\right] \label{eq:vshake_int_eom_pos}\\
    \dot{\mb{R}}(t+\dt) &= \dot{\mb{R}}(t) + \frac{\dt}{2}\mb{M}^{-1}\left[\mb{F}(t) + \gamma_{x}\mb{G}(t) + \mb{F}(t+\dt) + \gamma_{v}\mb{G}(t+\dt) \right] \label{eq:vshake_int_eom_vel} 
\end{align}
\end{subequations}
Here, $\dt$ is the time step, and $\gamma_{x}$ and $\gamma_{v}$ are the position and velocity Lagrange multipliers, respectively, needed to ensure that the trajectory remains in the diabatic seam space.
Details about solving for these Lagrange multipliers are provided in Section~\ref{sec:si_lagrange} of the SI. 
The Lagrange multipliers are determined to be converged when $\abs{\sigma(\mb{R})}<\tau$ is satisfied for some given error tolerance $\tau$ at each time step of the trajectory. 
Note that $\gp$, the reaction coordinate vector that is perpendicular to the seam and points from the reactant to the product (Figure~\ref{fig:seam_schematic}), is $-\mb{G}$ evaluated at the MECP, i.e.  $\gp = \nabla_{\mb{R}}\sigma(\mb{R}_{\text{MECP}})$.

All NEO calculations in this work were performed in a developer branch of Q-Chem 6.\cite{epifanovsky_software_2021} These calculations used the B3LYP\cite{lee_development_1988, becke_density-functional_1993} electronic exchange--correlation functional and the epc17-2\cite{brorsen_multicomponent_2017, yang_development_2017} electron--proton correlation functional with the aug-cc-pVDZ \cite{dunning_gaussian_1989} electronic basis set and the PB4-D \cite{yu_development_2020} protonic basis set. 
The exponents of the PB4-D basis for the D isotope were scaled by $\sqrt{2}$ to account for the mass difference.
Note that this level of theory is lower than the level used in ref~\citenum{paenurk_nuclearelectronic_2026}, and we do not anticipate the couplings computed herein to be as quantitatively accurate as those computed in that work. 
Our goal is to present a proof-of-concept for the utility of the V-SHAKE approach. 
The current implementation of V-SHAKE samples geometries in which the tunneling occurs between the diabatic NEO-DFT ground vibronic states.
Generalization of the V-SHAKE approach to sample geometries in which the tunneling occurs between arbitrary diabatic states (e.g., between the ground reactant vibronic state and the first-excited product vibronic state) is straightforward but is beyond the scope of this work. 

All V-SHAKE trajectories were initialized from the NEO-MECP found according to the procedure described in ref~\citenum{paenurk_nuclearelectronic_2026}.
The trajectories were thermostatted to either 300~K or 1000~K using the stochastic modified Berendsen method\cite{bussi_canonical_2007} with a time constant of 10 fs.
A total of 12 and 16 initial velocities were sampled for the CBO and HEO molecular systems, respectively, for each isotope.
The initial velocities were chosen according to standard Maxwell--Boltzmann sampling at the specified temperature.
Half of the initial velocities for each system were sampled at 300~K, with the other half sampled at 1000~K.
All trajectories were thermostatted to the temperatures at which their initial velocities were generated.
The use of these two different temperatures ensured that regions both near and far from the MECP were adequately sampled. 
All trajectories utilized a time step of 1.0 fs.
The trajectories were propagated for long enough to obtain a total of at least 1000 and 2000 configurations for CBO and HEO, respectively, for each isotope. 
The convergence criterion for the Lagrange multipliers $\gamma_x$ and $\gamma_v$ was $\tau=10^{-6}$ a.u. The evaluation of the NEO-DFT energies needed during the root search routine for $\gamma_x$ and $\gamma_v$ occurred only after all nuclear basis function centers were variationally optimized to their appropriate minima. 
If a configuration corresponding to a single-well Born--Oppenheimer potential energy surface was encountered, a trajectory recovery method was used to generate a new search direction within the seam. 
See Sections~\ref{sec:si_bl_msdft} and \ref{subsec:si_recovery} of the SI for more details about these procedures.

\begin{figure}
    \centering 
    \includegraphics[width=3.25in]{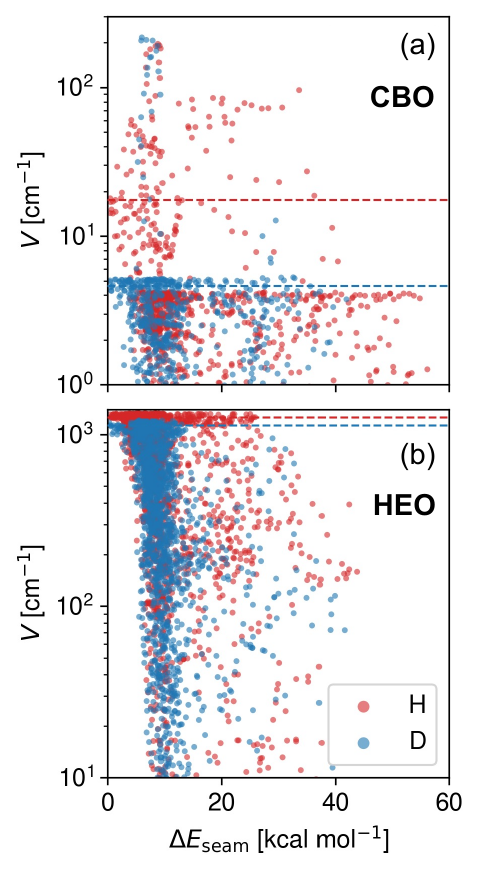}
    \caption{
    Scatter plots of vibronic couplings versus energies relative to their respective NEO-MECPs sampled with V-SHAKE for (a) CBO and (b) HEO.
    The red and blue horizontal dashed lines represent the value of the vibronic coupling at the corresponding MECP. 
    The vibronic couplings remained below the value at the corresponding MECP for the structures sampled for HEO, but higher vibronic couplings were sampled for CBO. 
    Only results for structures with vibronic couplings greater than 1 cm$^{-1}$ and 10 cm$^{-1}$ are shown for CBO and HEO, respectively.
    } 
    \label{fig:vib_coup_e_seam}
\end{figure}

We begin by analyzing the distribution of sampled vibronic couplings against their corresponding energies relative to the MECP for both CBO and HEO. 
Because all structures in the diabatic seam satisfy $E\ur=E\up$, where $V=\Delta/2$, the distribution of vibronic couplings is equivalent to the distribution of tunneling splittings.
Figure~\ref{fig:vib_coup_e_seam} shows that the diabatic seam spaces for CBO and HEO for both H and D isotopes exhibit a large spread of vibronic couplings over a wide range of energies relative to their MECPs, with sampled structures having relative energies as high as 40 or 60 kcal mol$^{-1}$ in HEO and CBO, respectively. 
Here, we report only the results for structures with vibronic couplings greater than 1 or 10 cm$^{-1}$ for CBO and HEO, respectively. 
As discussed in the SI of ref~\citenum{paenurk_nuclearelectronic_2026}, smaller values of the vibronic coupling computed with NEO-MSDFT are quantitatively unreliable because of numerical noise in regions of low protonic or deuteronic density. 
Such issues also occur in other fragment-based NOCI schemes, such as constrained DFT configuration interaction.\cite{mavros_communication_2015}

For HEO, there is a general tendency for the vibronic coupling to decrease as the relative energy increases.
This behavior is consistent with the notion that higher-energy structures on the seam usually correspond to larger donor--acceptor distances, which are typically associated with smaller couplings. 
For both isotopes, very few structures are sampled with vibronic couplings appreciably larger than the vibronic coupling at the MECP.
The vibronic coupling decreases rapidly with increasing energy close to the MECP energy for both isotopes. 
Typically, the rate constant increases with the vibronic coupling. 
Thus, the structures with small vibronic couplings relative to the vibronic coupling at the MECP are not likely to play a significant role in the hydrogen transfer kinetics of HEO.
In this case, although the vibronic coupling does not remain constant for the energetically accessible structures, this violation of the Condon approximation does not impact the rate constant calculations. 

For CBO, an appreciable number of structures that are energetically close to their MECPs have markedly larger vibronic couplings for both isotopes, although this behavior is more pronounced for the H isotope. 
The sampling of structures with low relative energies but high vibronic couplings points to a larger degree of geometric freedom within the seam space for CBO than for HEO. 
Again, the Condon approximation is violated, but in this case, non-Condon effects may have an impact on the hydrogen transfer kinetics in CBO. 
As discussed below, further analysis indicates that the dominant coordinate influencing the vibronic coupling in the seam space is the donor--acceptor distance, $R\uda$. 
Thus, the Condon approximation is expected to be valid if the sampling within the seam space were performed with a fixed value of $R\uda$. 
In this case, theories that treat $R\uda$ explicitly and thermally average over this coordinate\cite{levich_theory_1970, borgis_curve_1996, soudackov_quantum_2005, paenurk_nuclearelectronic_2026} can reasonably apply the Condon approximation to the fundamental rate constant expression.

\begin{figure}
    \centering 
    \includegraphics[width=3.25in]{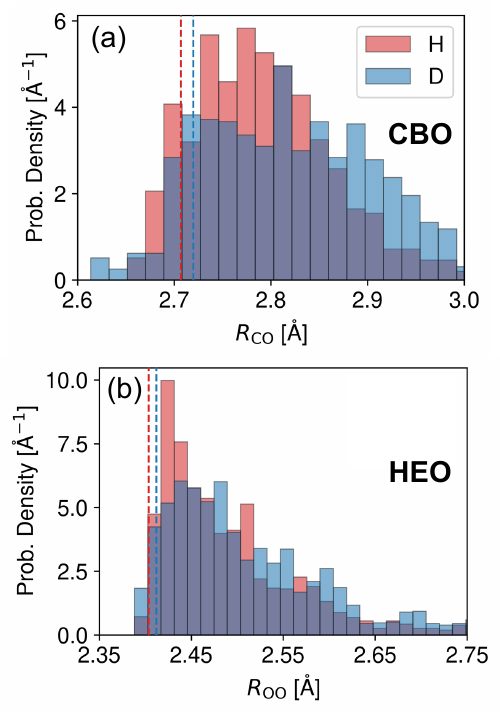}
    \caption{
    Normalized probability distributions of the donor--acceptor distances sampled with V-SHAKE for (a) CBO and (b) HEO. 
    The red and blue dashed vertical lines represent the value of the donor--acceptor distance at the corresponding MECP. 
    Both CBO and HEO sample relatively few structures with donor--acceptor distances shorter than those of their MECPs, although CBO samples more of such structures. 
    } 
    \label{fig:rda_dist}
\end{figure}

CBO and HEO exhibit similar trends in their distributions of sampled donor--acceptor distances, $R\uda$, as shown in Figure~\ref{fig:rda_dist}. 
H and D distributions for both systems indicate that structures with shorter values of $R\uda$ than those at their corresponding MECPs were not sampled very frequently. 
However, shorter donor--acceptor distances were sampled more frequently for CBO, explaining the greater number of structures with larger vibronic couplings than those at their MECPs (Figure~\ref{fig:vib_coup_e_seam}).
The lack of sampling of small $R\uda$ for HEO may be due to the rigidity of the molecule, but it may also be due to the diabatic states being ill-defined at small $R\uda$, where the transferring hydrogen moves on a single-well or nearly single-well Born--Oppenheimer potential energy surface, as discussed above.
Given the relatively short donor--acceptor distances at the MECPs for HEO, even shorter donor--acceptor distances are most likely not sampled frequently in HEO because the MECP is already very close to the single-well regime. 
Additionally, the donor--acceptor distance at the MECP is shorter for H than for D for both systems, consistent with stronger hydrogen-bonding interactions for the more delocalized H compared to D. \cite{reyes_investigation_2005, smith_isotope_2025}
Moreover, the $R\uda$ distribution is broader for D than for H in both systems, presumably due to slightly stronger hydrogen-bonding interactions for H.

\begin{figure}
    \centering 
    \includegraphics[width=3.25in]{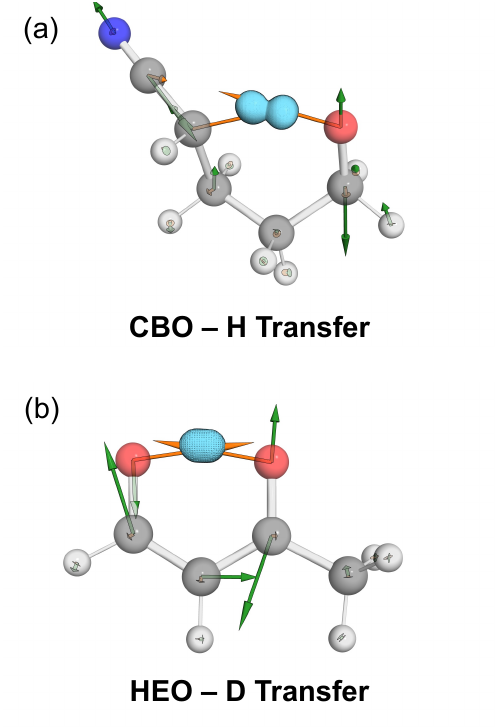}
    \caption{
    Visualization of the $\gp$ and $\gv$ vectors, shown in green and orange, respectively, for (a) H transfer in CBO and (b) D transfer in HEO. 
    The individual nuclear components of these vectors are overlaid on the classical nuclei for the structure of the corresponding NEO-MECP. 
    The $\gp$ vector for both systems primarily involves compression or elongation of the covalent bonds involving the donor and acceptor atoms, whereas the $\gv$ vectors primarily involve compression of the donor--acceptor distance.
    The NEO-MSDFT ground state quantum proton/deuteron densities are shown in cyan. 
    } 
    \label{fig:gperp_gradV}
\end{figure}

We now analyze the reaction coordinate, $\gp$ (Figure~\ref{fig:seam_schematic}), and the gradient of the vibronic coupling at the MECP, $\gv$, for H transfer in CBO and D transfer in HEO.
We provide the $\gp$ vector for D transfer in CBO, as well as the $\gv$ and $\gp$ vectors for H transfer in HEO, in Figure~\ref{fig:gperp_gradV_si} in the SI.
The $\gp$ and $\gv$ vectors are shown in green and orange, respectively, in Figure~\ref{fig:gperp_gradV}, displayed with relative lengths and directions overlayed on top of their respective nuclei. 
We have also rendered motion along these vectors as GIFs that are accessible on GitHub (see the statement on Data Availability below).

For D transfer in HEO, $\gp$ involves mainly (i) compression of the donor C$-$O bond and elongation of the acceptor C$-$O bond and (ii) rocking of the central carbon of the carbon backbone toward the acceptor.  
These motions are related to the redistribution of double bonds in the carbon backbone (Figure~\ref{fig:lewis_dens}a).
Such motions are known to break the symmetry of the double-well potential in which the transferring hydrogen moves, thereby breaking the degeneracy of the reactant and product localized NEO-DFT vibronic states.
This breaking of degeneracy is not achieved by $R\uda$ compression or elongation but rather by stabilizing the product state relative to the reactant state, causing localization of the proton/deuteron density in the acceptor well. 
The reaction coordinate for H transfer in CBO also exhibits similar behavior, with the C--O bond elongating to stabilize the product state and the carbon donor stretching towards the carbon of the CN group to destabilize the reactant state, accompanied by complementary C$\equiv$N elongation. 

As expected, for both systems shown in Figure~\ref{fig:gperp_gradV}, $\gv$ is dominated by donor--acceptor compression. 
We have found $\gp$ and $\gv$ to be approximately orthogonal to each other in these systems (see Section~\ref{sec:si_supp_plots} of the SI). 
This orthogonality indicates that the reaction coordinate for hydrogen tunneling is not dominated by the motion that most strongly impacts the vibronic coupling, but rather is dominated by the motion that breaks the degeneracy of the reactant and product diabatic vibronic states with minimal contributions from $R\uda$. 
This finding supports the assumption made in NEO-GRT and other rate theories that the reaction coordinate and $R\uda$ are uncoupled,\cite{paenurk_nuclearelectronic_2026} allowing a thermal averaging procedure to account for the dependence of $V$ on $R\uda$.

In diabatic seam space sampling, it may be useful to have a purely data-driven method for determining important molecular motions influencing the vibronic coupling.
Such a method would identify the coordinates that most strongly influence $V$ while preserving degeneracy between the diabatic states, without requiring any \textit{a priori} knowledge of the system.
We explored this possibility with sliced inverse regression (SIR),\cite{li_sliced_1991, bernard-michel_gaussian_2009} a dimensionality reduction technique that can identify such motions given only the structures and couplings of sampled points in the diabatic seam.
We performed the SIR routine on the sampled points for each system and provide the results and analysis in Section~\ref{sec:si_sir} of the SI. 
The SIR routine appropriately identified the primary coordinate within each seam that controls the vibronic coupling to be the donor--acceptor distance $R\uda$. 
For more complex systems, such an analysis may be able to identify other important coordinates that strongly impact the vibronic coupling within the seam. 
These additional coordinates may need to be treated explicitly within the NEO-GRT\cite{paenurk_nuclearelectronic_2026} formalism and other rate theories.

In this work, we have introduced the V-SHAKE approach to sample diabatic seams between reactant and product NEO-DFT diabatic vibronic surfaces in hydrogen tunneling systems.
We used this approach to analyze the seam spaces for the CBO and HEO compounds shown in Figure~\ref{fig:lewis_dens}a for both H and D isotopes.
We found that the Condon approximation, i.e., the assumption that the vibronic coupling does not vary significantly in the energetically accessible crossing region, does not hold rigorously for either system. 
We also introduced a general data-driven method for identifying motions within the seam that directly modulate the coupling. 
Application of this data-driven method, as well as analysis of the gradient of the vibronic coupling at the MECP, indicates that the donor--acceptor distance is the dominant coordinate influencing the vibronic coupling.
Analysis of the reaction coordinate $\gp$ at each MECP indicates that hydrogen tunneling is driven by compression and elongation of the covalent bonds involving the donor and acceptor atoms, thereby stabilizing the product state relative to the reactant state. 
The reaction coordinate and the gradient of the vibronic coupling at the MECP are found to be nearly orthogonal. 
These findings provide validation for several assumptions underlying NEO-GRT\cite{paenurk_nuclearelectronic_2026} and other hydrogen transfer theories that rely on thermal averaging over the donor--acceptor distance. 

The methods introduced herein provide a direct protocol for comprehensively sampling the seam spaces that govern hydrogen tunneling reactions. 
The information provided by such calculations provides insight into the fundamental physical principles underlying H and D tunneling processes and is relevant to calculations of rate constants and KIEs.
The generalization of the V-SHAKE approach to sample seam spaces between higher-lying diabatic vibronic states is straightforward but will require further methodological developments. 
The V-SHAKE approach can also be combined with condensed-phase NEO methods \cite{chow_nuclear-electronic_2023, chow_nuclear_2024} to elucidate hydrogen tunneling in complex chemical or biological systems.

\section*{Data Availability}
The data underlying this study are openly available on GitHub at \href{https://github.com/joseph-dickison20/V-SHAKE}{https://github.com/joseph-dickison20/V-SHAKE}.
 
\section*{Supporting Information}
Technical details on the formulation of NEO-MSDFT in this work; 
details on numerically solving for the Lagrange multipliers to propagate the V-SHAKE dynamics in eq~\eqref{eq:vshake_int_eom};
SIR analysis of V-SHAKE data;
supplemental plots for CBO and HEO systems.

\section*{Acknowledgments}
This work was supported by the National Science Foundation Grant No. CHE-2408934. 
The authors thank Andreas Ghosh and Jang Mok Yoo for their helpful discussions and comments on the manuscript.
The authors thank Dr. Scott Garner, Dr. Richard Kang, Nicholas Boyer, Tim Duong, Rowan Goudy, Jack Morgenstein, Logan Smith, and Millan Welman for useful discussions. 

\providecommand{\latin}[1]{#1}
\makeatletter
\providecommand{\doi}
  {\begingroup\let\do\@makeother\dospecials
  \catcode`\{=1 \catcode`\}=2 \doi@aux}
\providecommand{\doi@aux}[1]{\endgroup\texttt{#1}}
\makeatother
\providecommand*\mcitethebibliography{\thebibliography}
\csname @ifundefined\endcsname{endmcitethebibliography}
  {\let\endmcitethebibliography\endthebibliography}{}

\clearpage

\setcounter{equation}{0}
\setcounter{figure}{0}
\setcounter{table}{0}
\setcounter{section}{0}
\renewcommand{\theequation}{S\arabic{equation}}
\renewcommand{\thefigure}{S\arabic{figure}}
\renewcommand{\thetable}{S\arabic{table}}
\renewcommand{\thesection}{S\arabic{section}}
\renewcommand{\thesubsection}{S\arabic{section}.\arabic{subsection}}
\renewcommand{\theHequation}{SI.\arabic{equation}}
\renewcommand{\theHfigure}{SI.\arabic{figure}}
\renewcommand{\theHtable}{SI.\arabic{table}}
\renewcommand{\theHsection}{SI.\arabic{section}}
\renewcommand{\theHsubsection}{SI.\arabic{section}.\arabic{subsection}}
\renewcommand{\bibnumfmt}[1]{[S#1]}
\renewcommand{\citenumfont}[1]{S#1}
\renewcommand{\thepage}{S\arabic{page}}
\setcounter{page}{1}

\makeatletter
\def\mcitetrackID{si}
\def\mcitebibtrackID{si}
\makeatother

\begin{center}
\vspace*{0.5in}
{\sffamily\bfseries\LARGE Supporting Information:\\[0.3em]
Diabatic Seam Space Sampling for Hydrogen Tunneling Systems with Nuclear--Electronic Orbital Theory\par}
\vspace{1.2em}
{\sffamily\large Joseph A. Dickinson\orcidlink{0000-0002-5601-7050}, Eno Paenurk\orcidlink{0000-0002-6921-757X}, Sharon Hammes-Schiffer\orcidlink{0000-0002-3782-6995}\par}
\vspace{0.8em}
{\itshape\small Department of Chemistry, Yale University, New Haven, CT 06520, USA\\
Fakultät für Chemie und Pharmazie, Universität Regensburg, 93053 Regensburg, Germany\\
Department of Chemistry, Princeton University, Princeton, NJ 08544, USA\par}
\vspace{0.6em}
{\small shs566@princeton.edu\par}
\end{center}

\newpage

\tableofcontents

\newpage

\section{Block-Localized NEO-MSDFT}\label{sec:si_bl_msdft}

In this section, we provide technical details regarding the NEO-MSDFT implementation used in this work. 
Due to limitations of the epc functionals and the form of the off-diagonal Hamiltonian matrix elements, it has been necessary to apply a scaling factor\cite{si-yu_nuclear-electronic_2020,si-dickinson_generalized_2023} to the value of the overlap between NEO-DFT diabatic vibronic states in $\mathbf{S}$ to obtain accurate tunneling splittings compared to grid-based\cite{si-webb_fourier_2000} references. 
The implementation of the NEO-MSDFT procedure used in this work utilizes a block-localized formalism \cite{si-cembran_block-localized_2009} that was not utilized in our previous work on  NEO-MSDFT. 
In addition to being more computationally efficient, the use of the block-localized formalism is more consistent with the V-SHAKE approach and does not result in appreciably different proton densities or energies compared to the original implementation.

We provide information on the block-localization procedure in Section~\ref{subsec:si_bl_msdft}, the justification for its use in V-SHAKE in Section~\ref{subsec:si_justify_bl}, and details on the scaling scheme reparameterization in Section~\ref{subsec:si_overlap}.
In this section only, we will refer to the original NEO-MSDFT implementation simply as ``NEO-MSDFT'' and the block-localized form as ``NEO-BL-MSDFT.'' 
Note that in the main text, we make no such distinction, opting to refer to the block-localized formalism simply as ``NEO-MSDFT'' to avoid confusion.

\subsection{NEO-MSDFT vs NEO-BL-MSDFT}\label{subsec:si_bl_msdft}

For simplicity, we only consider the case of a single tunneling hydrogen and a two-state diabatic expansion (eq~\eqref{eq:msdft_expansion} of the main text).
In NEO-MSDFT, two nuclear basis function centers must be employed: one closer to the hydrogen donor with position $\mc{R}\ud$ and another closer to the hydrogen acceptor with position $\mc{R}\ua$.
Both $\mc{R}\ud$ and $\mc{R}\ua$ are usually found via variational optimization of each center independently of the other following the generation of a biased guess of the diabatic electronic and/or protonic densities. 
This procedure is described further in our previous work. \cite{si-yu_nuclear-electronic_2020,si-dickinson_generalized_2023} 

Once $\mc{R}\ud$ and $\mc{R}\ua$ have been determined, NEO self-consistent field (NEO-SCF)\cite{si-liu_simultaneous_2022,si-chow_efficient_2026} calculations are performed for each NEO-DFT diabatic vibronic state to obtain the nuclear and electronic densities used to construct the matrices of eq~\eqref{eq:msdft_eigenproblem} in the main text.
In the original NEO-MSDFT implementation, these diabatic NEO-SCF calculations were performed with nuclear and electronic basis functions centered at both $\mc{R}\ud$ and $\mc{R}\ua$, i.e., each diabatic vibronic state can have nonzero density at its ``opposing'' center, although such density is usually very small. 
In NEO-BL-MSDFT, however, these diabatic NEO-SCF calculations do not include the opposing center, i.e., for the reactant and product NEO-DFT diabatic vibronic states, there is strictly no contribution to the electronic and nuclear densities from the basis functions associated with the opposing center. 
Thus, in NEO-MSDFT, the NEO-DFT diabatic vibronic states are dependent on both $\mc{R}\ud$ and $\mc{R}\ua$, whereas in NEO-BL-MSDFT, the NEO-DFT diabatic vibronic states are only dependent on the corresponding dominant center, $\mc{R}\ud$ or $\mc{R}\ua$. 
As the block-localized form uses half the number of nuclear basis functions compared to the original implementation, it is generally more computationally efficient, providing a non-negligible increase in speed for the V-SHAKE simulations.

\subsection{Justification for Using NEO-BL-MSDFT in V-SHAKE}\label{subsec:si_justify_bl}

In general, block-localization is a small change that does not greatly impact the nuclear densities or the energies of the diabatic vibronic states. 
However, depending on the level of theory and DFT grid used, two diabatic states that were found to satisfy $\abs{\sigma(\mb{R})}<\tau$ when the opposing center was not present need not necessarily satisfy $\abs{\sigma(\mb{R})}<\tau$ when the opposing center is present, given orbital reoptimization in the full basis of $\mc{R}\ud$ and $\mc{R}\ua$.
Thus, in order to ensure that $\sigma(\mb{R})$ remains constant following the solution for $\gamma_x$ (see Section~\ref{subsec:si_sigma_continuity} below), NEO-BL-MSDFT is the more justifiable choice in V-SHAKE simulations over standard NEO-MSDFT. 
In Section~\ref{sec:si_lagrange}, we discuss the details of the search routine for $\gamma_x$ and provide more context for why NEO-BL-MSDFT is more appropriate for propagating V-SHAKE dynamics.

\subsection{Overlap Scaling Scheme}\label{subsec:si_overlap}

The original overlap scaling scheme was parametrized for diabatic vibronic states using the full basis centered on both $\mc{R}\ud$ and $\mc{R}\ua$. 
Here, we must reparameterize the scaling scheme for the block-localized formalism. 
We review this scaling scheme below.

As originally introduced in ref~\citenum{si-yu_nuclear-electronic_2020}, the value of the overlap, $S=\langle\tilde{\Psi}\ud\vert\tilde{\Psi}\ua\rangle$, is replaced with a scaled value of the overlap, $S^\prime$.
This $S^\prime$ replaces all instances of $S$ in the working equations of NEO-MSDFT\cite{si-yu_nuclear-electronic_2020,si-dickinson_generalized_2023} according to
\begin{equation}\label{eq:si_original_overlap_correction}
    S^\prime = \alpha S^\beta
\end{equation}
where $\alpha$ and $\beta$ are real constants. 
The values of $\alpha$ and $\beta$ were obtained by fitting NEO-MSDFT tunneling splittings to numerically exact grid-based tunneling splittings \cite{si-webb_fourier_2000} for the FHF$^{-}$ system at a range of F--F distances, $R_{\mathrm{FF}}$.
This fitting routine, although simple, was found to result in transferable parameters that allowed NEO-MSDFT to predict quantitatively accurate hydrogen tunneling splittings for a diverse array of single-proton\cite{si-yu_nuclear-electronic_2020} and multiple-proton\cite{si-dickinson_generalized_2023} transfer systems. 

Following a similar FHF$^{-}$ fitting routine, we computed the hydrogen tunneling splittings using NEO-BL-MSDFT at the B3LYP\cite{si-lee_development_1988,si-becke_density-functional_1993}/epc17-2\cite{si-brorsen_multicomponent_2017,si-yang_development_2017} level of theory with the aug-cc-pVDZ electronic basis set for the F atoms and the aug-cc-pV5Z\cite{si-dunning_gaussian_1989} electronic basis set for the H/D atom, as well as the PB4-D protonic basis set. \cite{si-yu_development_2020}
All exponents of the PB4-D protonic basis set for the D isotope were scaled by $\sqrt{2}$ compared to their values in ref~\citenum{si-yu_development_2020} to account for the mass difference.
Reference results were obtained using the procedure outlined in ref~\citenum{si-webb_fourier_2000} with a grid of $32^3$ points evaluated at the B3LYP/aug-cc-pVDZ level.
The fit was performed using an objective function balancing absolute and percentage error between NEO-BL-MSDFT and reference splittings, with the absolute error being weighted more heavily than the percentage error.
In contrast to previous fitting strategies, the overlap parameterizations for H and D were coupled via the constraint of having to share the same $\alpha$. 
This choice was made to avoid the paradoxical situation of H having a larger $S$ than D prior to scaling but a smaller $S^{\prime}$ after scaling for certain geometries. 
All code pertaining to this fitting routine, along with all NEO-BL-MSDFT input and output files, are openly available on GitHub (see the statement on Data Availability in the main text). 
The results of this fit are given below in Table~\ref{tab:si_fhf_overlap_fit}.

\begin{threeparttable}
\captionof{table}{\justifying NEO-BL-MSDFT Tunneling Splittings for FHF$^{-}$ Versus Reference Splittings,\tnote{\textit{a}} with the Optimized Values of $\alpha$ and $\beta$ Provided Next to Each Isotope}
\label{tab:si_fhf_overlap_fit}
\centering
\begin{tabularx}{\textwidth}{c|*{2}{>{\centering\arraybackslash}X}|*{2}{>{\centering\arraybackslash}X}}
\hline
 & \multicolumn{2}{c|}{H ($\alpha=0.56,\beta=1.10$)} & \multicolumn{2}{c}{D ($\alpha=0.56,\beta=0.98$)} \\
$R_{\mathrm{FF}}$ (\AA) & \mbox{NEO-BL-MSDFT} & Ref. & \mbox{NEO-BL-MSDFT} & Ref. \\
\hline
2.60 & 271.56 & 289.90 & 96.61 & 97.81 \\
2.65 & N/A\tnote{\textit{b}} & 144.03 & 16.03 & 30.37 \\
2.70 & 41.21 & 57.50 & 12.59 & 7.01 \\
2.75 & 21.92 & 18.91 & 5.84 & 1.29 \\
2.80 & 8.56 & 5.34 & 2.56 & 0.20 \\
\hline
\end{tabularx}
\begin{tablenotes}
\begin{spacing}{1.0}
\item[\textit{a}]\footnotesize All tunneling splittings in cm$^{-1}$.
\item[\textit{b}]\footnotesize The H isotope at $R_{\mathrm{FF}}=2.65$~\AA\ exhibited unstable NEO-DFT basis-center optimization and SCF convergence behavior; thus, it was excluded from the fitting routine. Such behavior can be rectified by using constrained NEO\cite{si-xu_constrained_2020} to define the diabatic states, as the constraints typically lead to more stable SCF solutions. Such an approach was used in ref~\citenum{si-paenurk_nuclearelectronic_2026} but not in the block-localized form discussed above. 
\end{spacing}
\end{tablenotes}
\end{threeparttable}
\bigskip

\FloatBarrier

In ref~\citenum{si-paenurk_nuclearelectronic_2026}, Paenurk et al.\ introduced a new scaling scheme that is generally more robust than the original scheme that is used herein. 
The form of this new scheme ensures that $S^\prime\rightarrow 1$ as $S\rightarrow1$.
In the original overlap scheme, $S^\prime\rightarrow\alpha$ as $S\rightarrow1$, which becomes less accurate for geometries with large overlap.
However, the use of the new scaling scheme in conjunction with V-SHAKE was found to consistently cause single-well geometries to be sampled, thereby limiting V-SHAKE from exploring regions of the diabatic seam space far from the NEO-MECP. 
The use of the original scheme in this work allows V-SHAKE to sample geometries far from the NEO-MECP on the seam. 
Additionally, the value of $\alpha$ found in the fitting procedure described herein is appreciably larger than previous $\alpha$ values, avoiding substantial damping of the overlap.

We emphasize that the V-SHAKE approach is generally independent of NEO-MSDFT. 
Within V-SHAKE, NEO-MSDFT is simply used as a means of computing vibronic couplings and tunneling splittings for the sampled configurations in the diabatic seam and propagating the classical nuclear dynamics.
In principle, V-SHAKE could also be formulated in conjunction with a more robust NEO wavefunction method, such as NEO complete active space SCF (NEO-CASSCF) or NEO multireference configuration interaction (NEO-MRCI).\cite{si-malbon_nuclear-electronic_2025,si-stein_computing_2025}

\section{Solving for Lagrange Multipliers During V-SHAKE}\label{sec:si_lagrange}

In this section, we provide details on the numerical solution of the Lagrange multipliers $\gamma_x$ and $\gamma_v$ introduced in eq~\eqref{eq:vshake_int_eom} of the main text. 
This routine must be performed at each step of a V-SHAKE trajectory to ensure that all trajectories remain on the diabatic seam.
We discuss this routine in Section~\ref{subsec:si_lagrange}, provide comments on the continuity of $\sigma$ in Section~\ref{subsec:si_sigma_continuity}, and describe a trajectory recovery routine when V-SHAKE samples single-well regimes in Section~\ref{subsec:si_recovery}.

\subsection{Lagrange Multiplier Optimization}\label{subsec:si_lagrange}

Cofer-Shabica et al.~\cite{si-cofer-shabica_marcus_2026} showed that $\gamma_x$ can be obtained via a one-dimensional root search solving
\begin{equation}\label{eq:si_gamma_x}
    \sigma\left(\tilde{\mb{R}}(t+\dt)+\gamma_x\frac{\dt^2}{2}\mb{M}^{-1}\mb{G}(t)\right) = 0
\end{equation}
where $\tilde{\mb{R}}(t+\dt)$ is the configuration that would be obtained at time $t+\dt$ via the velocity-Verlet algorithm\cite{si-swope_computer_1982} under no constraint at the current time step.
They also showed that once $\gamma_x$ was obtained numerically, $\gamma_v$ can be found analytically via
\begin{equation}\label{eq:si_gamma_v}
    \gamma_v = -\frac{2\tilde{\dot{\mb{R}}}(t+\dt)\mb{G}(t+\dt)}{\dt\mb{G}^{\text{T}}(t+\dt)\mb{M}^{-1}\mb{G}(t+\dt)}
\end{equation}
where $\tilde{\dot{\mb{R}}}(t+\dt)$ is defined as the partially unconstrained velocity given by
\begin{equation}\label{eq:si_unconstrained_vel}
    \tilde{\dot{\mb{R}}}(t+\dt) = \mb{Q} + \frac{\dt}{2}\mb{M}^{-1}\mb{F}(t+\dt)
\end{equation}
and the matrix $\mb{Q}$ is dependent on $\gamma_x$ according to
\begin{equation}
    \mb{Q} = \dot{\mb{R}}(t) + \frac{\dt}{2}\mb{M}^{-1}\left[\mb{F}(t)+\gamma_x\mb{G}(t)\right]
\end{equation}

Following Cofer-Shabica et al., we used bisection with the assumption of linearity to solve for the $\gamma_x$ that satisfies eq~\eqref{eq:si_gamma_x}, with a single Newton--Raphson step to bracket the initial guess. 
Each iteration of the root-finding procedure involves an evaluation of $\sigma$ at some $\mb{R}^*(\gamma^*_{x})$, defined as the the argument of the $\sigma$ function in eq~\eqref{eq:si_gamma_x} for some guessed $\gamma^*_{x}$
\begin{equation}\label{eq:si_r_trial}
    \mb{R}^*(\gamma^*_{x}) = \tilde{\mb{R}}(t+\dt)+\gamma^*_x\frac{\dt^2}{2}\mb{M}^{-1}\mb{G}(t)
\end{equation}
The procedure is declared complete once a $\gamma^*_{x}$ has been found satisfying $\abs{\sigma\left(\mb{R}^*(\gamma^*_x)\right)} < \tau$, where $\tau$ is a pre-defined error threshold between diabatic NEO-DFT energies.
In all V-SHAKE simulations in this work, we set $\tau = 10^{-6}$ a.u. 
Once such a $\gamma^*_x$ is found, it is the $\gamma_x$ used to propagate the equations-of-motion in eq~\eqref{eq:vshake_int_eom} of the main text, with $\gamma_v$ obtained via eq~\eqref{eq:si_gamma_v}.

\subsection{Continuity of the Diabatic Energy Difference}\label{subsec:si_sigma_continuity}

At each evaluation of $\sigma$, the basis function centers $\mc{R}\ud$ and $\mc{R}\ua$ defined in Section~\ref{sec:si_bl_msdft} are each optimized to their minimum on their respective NEO-DFT diabatic vibronic surfaces.
As discussed in Section~\ref{subsec:si_justify_bl}, these optimizations and energy evaluations occur for each diabat with only its corresponding basis function center present. 
Once $\gamma_x$ has been found, the most recent electronic and nuclear densities found for each diabat then enter the NEO-BL-MSDFT routine, where $\sigma$ is guaranteed to remain constant from the end of the root-finding procedure to the NEO-BL-MSDFT calculation because the densities are not reoptimized.
This property is not guaranteed with standard NEO-MSDFT, as reoptimizing each density via NEO-SCF calculations with both $\mc{R}\ud$ and $\mc{R}\ua$ present may break the $\abs{\sigma(\mb{R})}<\tau$ condition that was satisfied during the root-finding procedure for $\gamma_x$.

\subsection{Trajectory Recovery Routine}\label{subsec:si_recovery}

Eq~\eqref{eq:si_r_trial} shows that once $\tilde{\mb{R}}(t+\dt)$ is known, the vector $\dt^2/2\cdot\mb{M}^{-1}\mb{G}(t)$ defines the direction along which a new $\mb{R}^*$ is generated.
However, there is no guarantee that for any given $\tilde{\mb{R}}(t+\dt)$, there must be an $\mb{R}^*$ in the diabatic seam along $\dt^2/2\cdot\mb{M}^{-1}\mb{G}(t)$, i.e., the $\gamma_x$ search may result in $\mb{R}^*$ where the hydrogen moves in a single-well potential, detectable by $\mc{R}\ud$ and $\mc{R}\ua$ optimizing to the same position.
If this situation occurs at any point during a $\gamma_x$ search, to ensure that we continuously sample only double-well regions where hydrogen tunneling is allowed, we resample random Maxwell--Boltzmann velocities and scale them such that the kinetic energy of the system remains continuous.
These new velocities are then used to compute a new $\tilde{\mb{R}}(t+\dt)$ from which the $\gamma_x$ search can restart.
This process is repeated until a configuration on the diabatic seam is found and eq~\eqref{eq:si_gamma_x} is satisfied.
We emphasize that the V-SHAKE dynamics are not physically meaningful: V-SHAKE is simply a tool to comprehensively sample the diabatic seam.
Thus, strategies that alter the ``dynamics'' are not problematic, as long as the trajectory remains on the diabatic seam and the strategy does not bias sampling among the double-well configurations.

\section{Sliced Inverse Regression Analysis of V-SHAKE Data}\label{sec:si_sir}

In this section, we provide a purely data-driven methodology based on sliced inverse regression (SIR)\cite{si-li_sliced_1991,si-bernard-michel_gaussian_2009} to determine the coordinates over which the vibronic coupling varies most strongly within the seam. 
We refer the reader to refs~\citenum{si-li_sliced_1991} and~\citenum{si-bernard-michel_gaussian_2009} for the rigorous theory behind SIR and to the GitHub repository for the regression parameters and further computational details.
We present only the essential theory of SIR in the context of V-SHAKE here.

The seam space is formally one dimension less than the full configuration space, due to the single constraint that $E\ur=E\up$. 
However, it may be the case that only a small subset of these coordinates in the diabatic seam significantly impact the vibronic coupling.
We are interested in identifying this smaller subset of coordinates using the sampled structures in the diabatic seam and their associated vibronic couplings.

Assume that $V$ only effectively depends on $k$ coordinates, $\{\boldsymbol{\eta}_i\}_{i=1}^k\subset\mathbb{R}^{3N\uc}$, where $k\ll3N\uc$. 
SIR identifies these $\boldsymbol{\eta}_i$ after diagonalizing a matrix $\mb{A}$.
This $\mb{A}$ matrix is constructed following a ``whitening'' procedure for the structures (detailed elsewhere)\cite{si-li_sliced_1991,si-bernard-michel_gaussian_2009,si-gewers_principal_2021} according to
\begin{equation}\label{eq:si_amat_sir}
    \mb{A} = \sum_{j=1}^{N_{\text{s}}} \frac{n_j}{n}\bar{\mathbf{a}}_j\bar{\mathbf{a}}_j^{\text{T}}
\end{equation}
where $n$ is the total number of sampled points in the diabatic seam, $N_{\text{s}}$ is the chosen number of ``slices,'' and $n_j$ is the number of structures in the $j$-th ``slice.'' 
Each slice is defined as the set of all structures (following Kabsch-alignment\cite{si-kabsch_solution_1976} to the MECP) binned together after ordering them in ascending order of $V$ and populating each slice with a chosen number of structures per slice. 
As the desired number of structures per slice may not evenly divide the total number of sampled points $n$, one slice (here chosen to be the slice with the lowest-$V$ structures) does not have the predefined number of structures per slice.
Each $\bar{\mathbf{a}}_j\in\mathbb{R}^{3N\uc}$ in eq~\eqref{eq:si_amat_sir} is computed as the mean value of that (whitened) coordinate in the $j$-th slice.
The desired $\boldsymbol{\eta}_i$ are found after diagonalizing $\mb{A}$ and performing an ``un-whitening'' procedure on the raw eigenvectors to obtain physically meaningful coordinates.

For the sake of brevity, we only show $\boldsymbol{\eta}_1$, the primary coordinates over which $V$ is most sensitive within the seam, but other coordinates are plotted and animated on GitHub.
These $\boldsymbol{\eta}_1$ are shown in Figure~\ref{fig:sir} below. 
As expected, the dominant motion in each $\boldsymbol{\eta}_1$ is the donor--acceptor compression. 
We emphasize that SIR returns the directions within the seam from the MECP that change $V$ the most strongly. 
Performing this analysis only requires the sampled structures in the diabatic seam and their couplings, and it is generally applicable to all diabatic seam space sampling routines.\cite{si-cofer-shabica_marcus_2026}
For more complex systems than those studied herein, such an analysis can identify important in-seam coordinates that should be treated explicitly in rate theories.

\begin{figure}
    \centering 
    \includegraphics[width=6.5in]{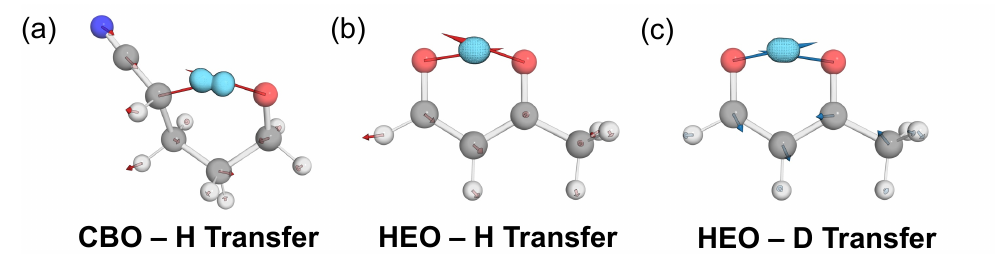}
    \caption{
    Visualization of the $\boldsymbol{\eta}_1$ vector, the primary in-seam motion that influences the vibronic coupling as found by SIR, for (a) H transfer in CBO, (b) H transfer in HEO, and (c) D transfer in HEO. 
    The individual nuclear components of these vectors are overlaid on the nuclei for the classical nuclear structure of the corresponding NEO-MECP. 
    These $\boldsymbol{\eta}_1$ vectors are mainly composed of donor--acceptor compression, as expected for these systems. 
    The NEO-MSDFT ground state quantum proton/deuteron densities are shown in cyan.
    The $\boldsymbol{\eta}_1$ vectors are colored according to their isotope (i.e., red for H and blue for D). 
    } 
    \label{fig:sir}
\end{figure}

\FloatBarrier
\section{Supplemental Analysis: $V$ vs $R\uda$ \& $\gp$ Orthogonality}\label{sec:si_supp_plots}

\begin{figure}
    \centering 
    \includegraphics[width=6.5in]{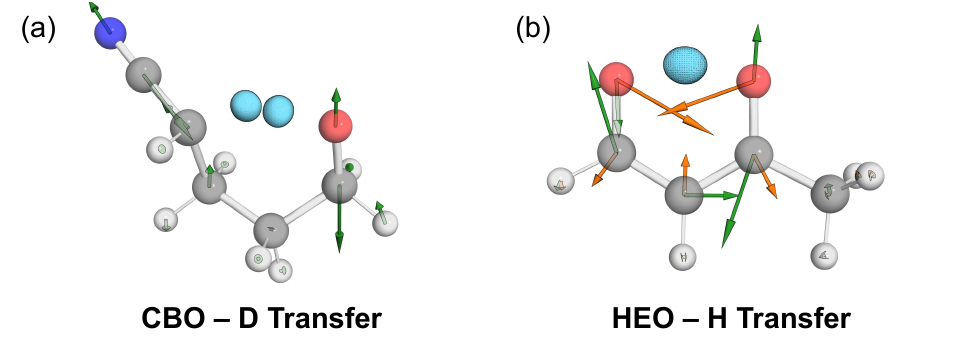}
    \caption{
    Visualization of (a) the $\gp$ vector, shown in green, for D transfer in CBO and (b) the $\gp$ and $\gv$ vectors, shown in green and orange, respectively, for H transfer in HEO. 
    The individual nuclear components of these vectors are overlaid on the classical nuclei for the structure of the corresponding NEO-MECP. 
    The $\gp$ vectors primarily involve compression or elongation of the covalent bonds involving the donor and acceptor atoms, whereas the $\gv$ vector for H transfer in HEO primarily involves compression of the donor--acceptor distance.
    The NEO-MSDFT ground state quantum proton/deuteron densities are shown in cyan.
    } 
    \label{fig:gperp_gradV_si}
\end{figure}

In this section, we provide further analyses of the vibronic couplings and reaction coordinates.
Figure~\ref{fig:gperp_gradV_si} shows the $\gv$ and $\gp$ vectors for H transfer in HEO. 
These vectors exhibit the same main characteristics as the corresponding D transfer vectors discussed in the main text (Figure~\ref{fig:gperp_gradV}b). 
Figure~\ref{fig:gperp_gradV_si} also shows the $\gp$ vector for D transfer in CBO. This vector exhibits the same main characteristics as the corresponding H transfer vector discussed in the main text (Figure~\ref{fig:gperp_gradV}a). 
The gradient of the vibronic coupling at the MECP is not shown for D transfer in CBO because the vibronic coupling at the MECP is $\sim 5$ cm$^{-1}$, which is in the regime where NEO-MSDFT vibronic couplings are less quantitatively reliable.

\FloatBarrier
\subsection{$V$ vs $R\uda$}\label{subsec:si_v_vs_rda}

\begin{figure}
    \centering 
    \includegraphics[width=3.33in]{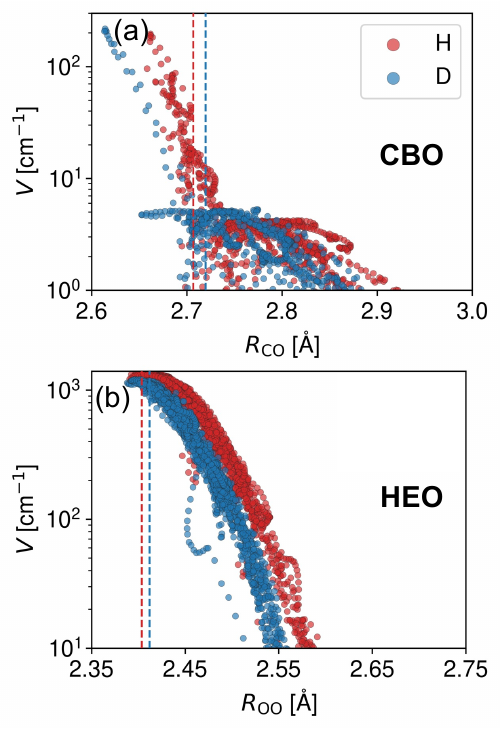}
    \caption{
    Vibronic couplings versus their donor--acceptor distances sampled with V-SHAKE for (a) CBO and (b) HEO. 
    The red and blue dashed vertical lines represent the value of the donor--acceptor distance at the corresponding MECP. 
    In general, the expected trend of approximately exponential decay of $V$ with $R\uda$ is observed for all systems in the regime corresponding to numerically reliable vibronic couplings (i.e., greater than $\sim 5$ cm$^{-1}$).
    Only results for structures with vibronic couplings greater than 1 cm$^{-1}$ and 10 cm$^{-1}$ are shown for CBO and HEO, respectively.
    } 
    \label{fig:si_v_vs_rda}
\end{figure}

Figure~\ref{fig:si_v_vs_rda} shows the vibronic couplings versus the corresponding donor--acceptor distances. 
In general, the expected trend of the vibronic coupling decaying approximately exponentially with increasing $R\uda$ is observed.
However, there are noticeable outliers to this trend. 
Although some outliers correspond to relatively high vibronic couplings, most of these outliers correspond to relatively small vibronic couplings. 
As discussed in the main text, NEO-MSDFT does not provide numerically reliable vibronic couplings smaller than $\sim 5$ cm$^{-1}$. Thus, the approximately exponential decay is not expected to be observed for the smaller vibronic couplings.

\FloatBarrier
\subsection{Orthogonality Analysis: $\boldsymbol{\eta}_1$ vs $\gp$}\label{subsec:si_orthog_sir}

\begin{figure}
    \centering 
    \includegraphics[width=6in]{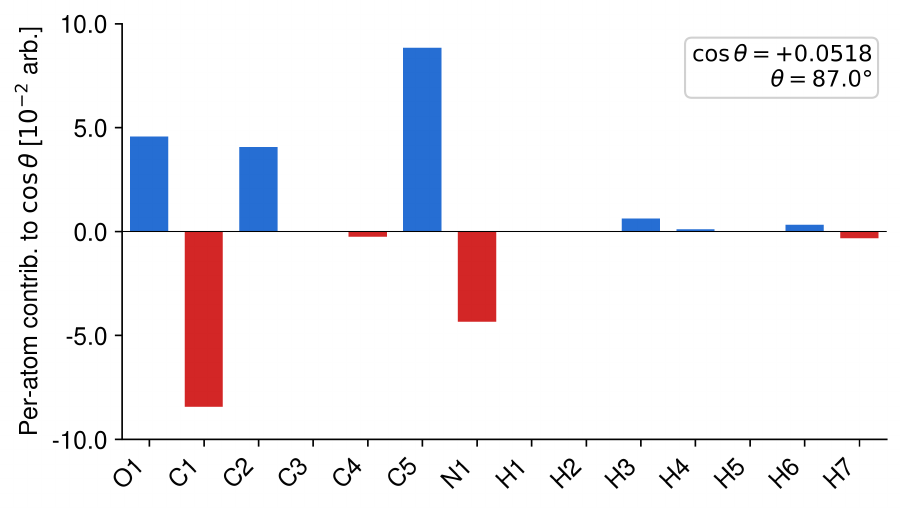}
    \caption{
    Individual per-atom contributions to the value of $\cos\theta$ between $\boldsymbol{\eta}_1$ and $\gp$ for H transfer in CBO.
    The value of $\cos\theta$ is provided in the top right, indicating that $\boldsymbol{\eta}_1$ is nearly orthogonal to $\gp$. 
    The color of the bars represents the sign of the contribution (i.e., blue for positive contributions and red for negative contributions). 
    } 
    \label{fig:cbo_sir_v_gperp}
\end{figure}

\begin{figure}
    \centering 
    \includegraphics[width=6in]{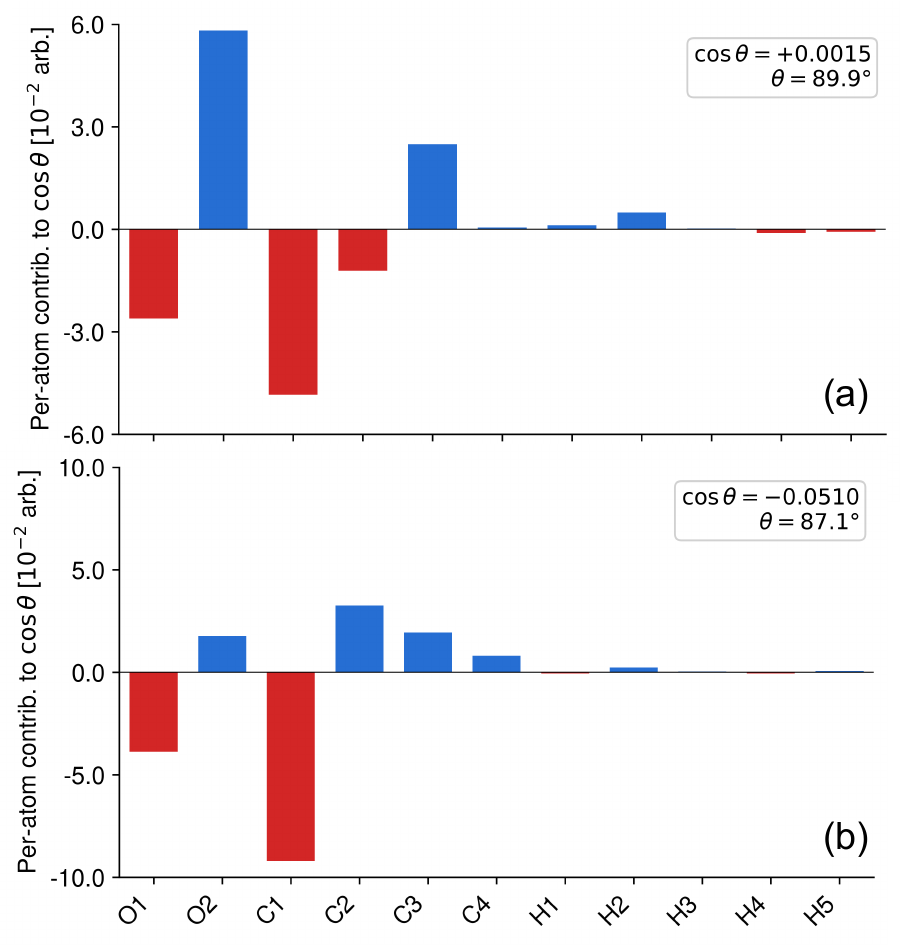}
    \caption{
    Individual per-atom contributions to the value of $\cos\theta$ between $\boldsymbol{\eta}_1$ and $\gp$ for (a) H transfer and (b) D transfer in HEO.
    The value of $\cos\theta$ is provided in the top right of each panel.
    Both angles are $\approx90^{\circ}$, indicating that $\boldsymbol{\eta}_1$ is nearly orthogonal to $\gp$. 
    The color of the bars represents the sign of the contribution (i.e., blue for positive contributions and red for negative contributions). 
    } 
    \label{fig:heo_sir_v_gperp}
\end{figure}

In this section, we provide the per-atom contribution to the cosine similarity between the $\boldsymbol{\eta}_1$ of each system and its associated reaction coordinate $\gp$. 
Note that D transfer in CBO was not analyzed due to its relatively small vibronic couplings.
We anticipate the $\boldsymbol{\eta}_1$ of each system to be orthogonal to its associated $\gp$, as $\boldsymbol{\eta}_1$ is the primary in-seam motion controlling $V$, whereas $\gp$ is explicitly orthogonal to the seam.
We observe this anticipated trend in all systems analyzed. 
The atomic labels used for the geometries provided in Tables~\ref{tab:si_heo_prot_mcep} through~\ref{tab:si_cbo_deut_mcep} in Section~\ref{sec:si_cart} are consistent with those used in Figures~\ref{fig:cbo_sir_v_gperp} and~\ref{fig:heo_sir_v_gperp}.
 
\FloatBarrier
\subsection{Orthogonality Analysis: $\gv$ vs $\gp$}\label{subsec:si_orthog_gradV}

\begin{figure}
    \centering 
    \includegraphics[width=6in]{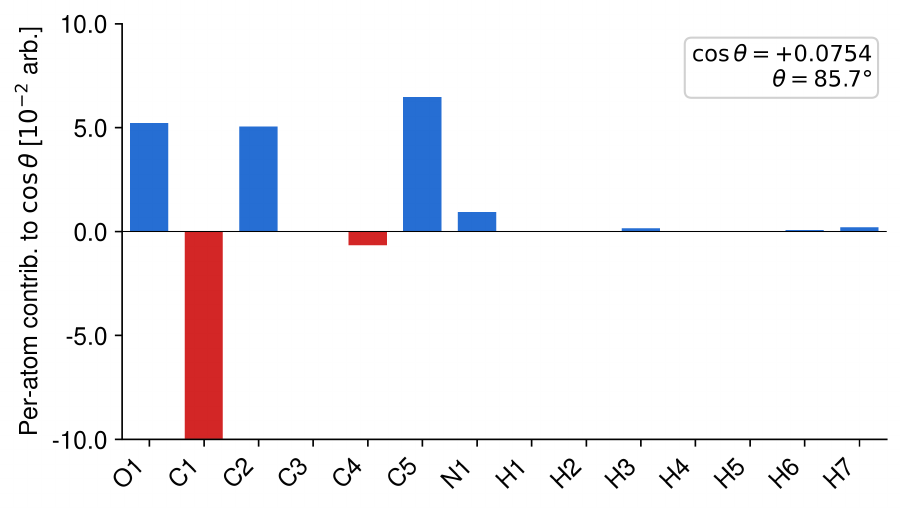}
    \caption{
    Individual per-atom contributions to the value of $\cos\theta$ between $\gv$ and $\gp$ for H transfer in CBO.
    The value of $\cos\theta$ is provided in the top right, indicating that $\gv$ is nearly orthogonal to $\gp$. 
    The color of the bars represents the sign of the contribution (i.e., blue for positive contributions and red for negative contributions).
    } 
    \label{fig:cbo_gradv_v_gperp}
\end{figure}

\begin{figure}
    \centering 
    \includegraphics[width=6in]{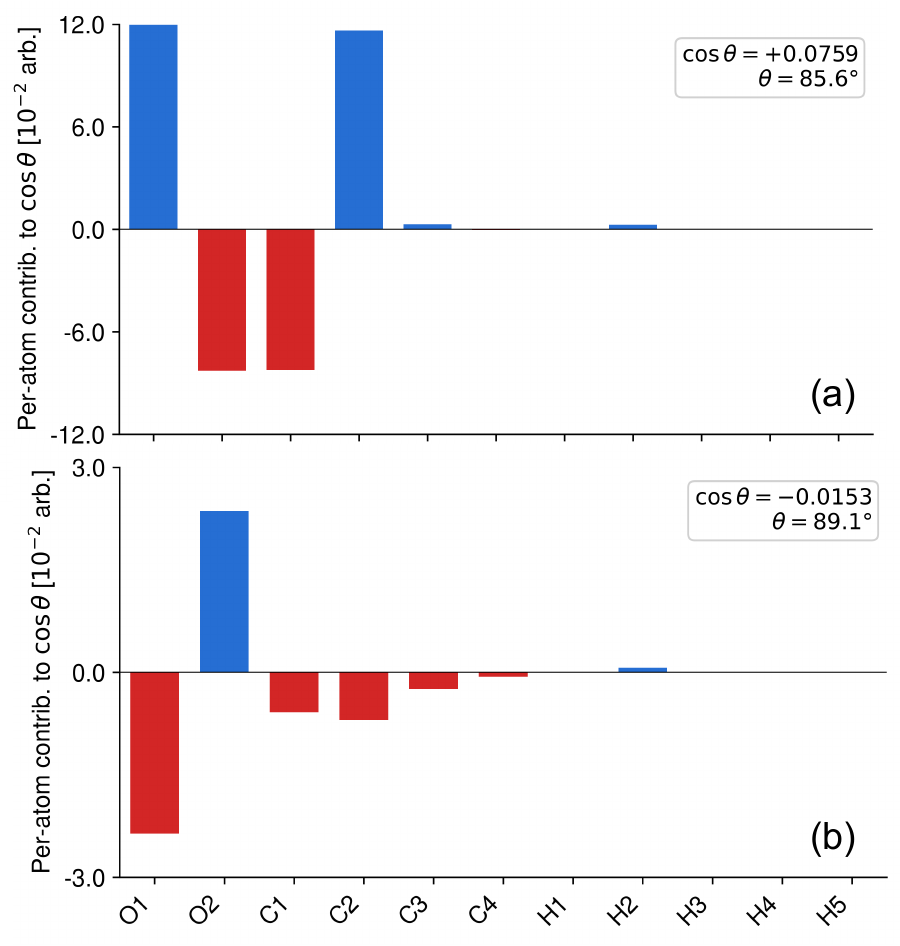}
    \caption{
    Individual per-atom contributions to the value of $\cos\theta$ between $\gv$ and $\gp$ for (a) H transfer and (b) D transfer in HEO.
    The value of $\cos\theta$ is provided in the top right, indicating that $\gv$ is nearly orthogonal to $\gp$. 
    The color of the bars represents the sign of the contribution (i.e., blue for positive contributions and red for negative contributions).
    } 
    \label{fig:heo_gradv_v_gperp}
\end{figure}

In this section, we provide the per-atom contribution to the cosine similarity between the $\gv$ of each system and its associated reaction coordinate $\gp$. Note that D transfer in CBO was not analyzed due to its relatively small vibronic couplings.
The atomic labels used for the geometries provided in Tables~\ref{tab:si_heo_prot_mcep} through~\ref{tab:si_cbo_deut_mcep} in Section~\ref{sec:si_cart} are consistent with those used in Figures~\ref{fig:cbo_gradv_v_gperp} and~\ref{fig:heo_gradv_v_gperp}.
We found $\gv$ and $\gp$ to be orthogonal to each other for all systems analyzed. 
This finding implies that the motions promoting the overall reaction from reactant to product are independent of the motions that most strongly influence the vibronic coupling. 

\newpage
\section{Cartesian Coordinates}\label{sec:si_cart}

In this section, we provide the Cartesian coordinates of the NEO-MECPs from which all V-SHAKE trajectories were initialized. 
These NEO-MECPs were found at the B3LYP/aug-cc-pVDZ epc17-2/PB4-D level of theory.
In the tables below, ``L$_{\text{D}}$'' and ``L$_{\text{A}}$'' represent the optimized positions of the donor and acceptor nuclear basis function centers for these MECPs, where ``L'' is either H or D depending on the isotope. 
The only isotopically exchanged nucleus in each system is the tunneling hydrogen. 

\vfill

\centering
\begin{threeparttable}
\captionof{table}{\justifying NEO-MECP for H transfer in HEO\tnote{\textit{a}}}
\begin{tabular}{c c c c}\label{tab:si_heo_prot_mcep}
Atom & X & Y & Z \\
\hline
O$_1$ & -1.1910317116 & \phantom{-}0.8506806972 & \phantom{-}0.0000000110 \\
O$_2$ & \phantom{-}1.2124106526 & \phantom{-}0.8152615939 & -0.0000000293 \\
C$_1$ & -1.1879682023 & -0.4350558664 & \phantom{-}0.0000000082 \\
C$_2$ & \phantom{-}1.2170405434 & -0.4704082827 & -0.0000000344 \\
C$_3$ & -0.0050176819 & -1.1802123172 & -0.0000000097 \\
C$_4$ & \phantom{-}2.5525396594 & -1.1567934264 & -0.0000000910 \\
H$_1$ & -0.0317410175 & -2.2661796990 & -0.0000000091 \\
H$_2$ & -2.1684127467 & -0.9323924939 & \phantom{-}0.0000000203 \\
H$_3$ & \phantom{-}3.1230929542 & -0.8411435131 & \phantom{-}0.8847715179 \\
H$_4$ & \phantom{-}2.4507176319 & -2.2475562919 & \phantom{-}0.0000001377 \\
H$_5$ & \phantom{-}3.1230926750 & -0.8411438823 & -0.8847720153 \\
H\ud & -0.0759427294 & \phantom{-}1.0929120598 & -0.0000000060 \\
H\ua & \phantom{-}0.1016372367 & \phantom{-}1.0862772158 & -0.0000000085 \\
\end{tabular}
\begin{tablenotes}
\item[\textit{a}]\footnotesize All values are in \AA.
\end{tablenotes}
\end{threeparttable}

\vfill
\clearpage

\null
\vfill

\centering
\begin{threeparttable}
\captionof{table}{\justifying NEO-MECP for D transfer in HEO\tnote{\textit{a}}}
\begin{tabular}{c c c c}\label{tab:si_heo_deut_mcep}
Atom & X & Y & Z \\
\hline
O$_1$ & \phantom{-}1.1949655617 & \phantom{-}0.8506201574 & -0.0000000109 \\
O$_2$ & -1.2167587547 & \phantom{-}0.8140536262 & \phantom{-}0.0000000282 \\
C$_1$ & \phantom{-}1.1888516183 & -0.4350353487 & -0.0000000080 \\
C$_2$ & -1.2177014165 & -0.4716057039 & \phantom{-}0.0000000333 \\
C$_3$ & \phantom{-}0.0054198500 & -1.1793195732 & \phantom{-}0.0000000089 \\
C$_4$ & -2.5517851987 & -1.1608726015 & \phantom{-}0.0000000907 \\
H$_1$ & \phantom{-}0.0328021370 & -2.2653114688 & \phantom{-}0.0000000095 \\
H$_2$ & \phantom{-}2.1685176070 & -0.9339057680 & -0.0000000185 \\
H$_3$ & -3.1230641618 & -0.8464989877 & -0.8847560050 \\
H$_4$ & -2.4477196602 & -2.2514189184 & -0.0000001390 \\
H$_5$ & -3.1230638778 & -0.8464993578 & \phantom{-}0.8847565049 \\
D\ud & \phantom{-}0.1120941500 & \phantom{-}1.0987950938 & \phantom{-}0.0000000049 \\
D\ua & -0.1377133642 & \phantom{-}1.0890113804 & \phantom{-}0.0000000092 \\
\end{tabular}
\begin{tablenotes}
\item[\textit{a}] \footnotesize All values are in \AA.
\end{tablenotes}
\end{threeparttable}

\vfill
\clearpage

\null
\vfill

\centering
\begin{threeparttable}
\captionof{table}{\justifying NEO-MECP for H transfer in CBO\tnote{\textit{a}}}
\begin{tabular}{c c c c}\label{tab:si_cbo_prot_mcep}
Atom & X & Y & Z \\
\hline
O$_1$ & \phantom{-}0.3673274496 & -0.0093158500 & \phantom{-}2.5662143167 \\
C$_1$ & -0.1107906321 & \phantom{-}0.0037302670 & -0.0980784527 \\
C$_2$ & -0.4608838577 & \phantom{-}1.0723528938 & \phantom{-}2.7277759956 \\
C$_3$ & -0.6250722138 & \phantom{-}1.9346357927 & \phantom{-}1.4451455904 \\
C$_4$ & -1.1327217584 & \phantom{-}1.1131260569 & \phantom{-}0.2512911127 \\
C$_5$ & -0.6194440043 & -1.0245988260 & -0.9746105291 \\
N$_1$ & -1.0338644636 & -1.8776006849 & -1.6599202563 \\
H$_1$ & \phantom{-}0.3526578547 & \phantom{-}2.3822592281 & \phantom{-}1.1914667075 \\
H$_2$ & -1.3256826071 & \phantom{-}2.7658887806 & \phantom{-}1.6500889353 \\
H$_3$ & -1.3473591661 & \phantom{-}1.7615964690 & -0.6175246526 \\
H$_4$ & -2.0868842690 & \phantom{-}0.6371474954 & \phantom{-}0.5319281059 \\
H$_5$ & \phantom{-}0.7969716537 & \phantom{-}0.4368221322 & -0.5467987550 \\
H$_6$ & -1.5074253115 & \phantom{-}0.7828565855 & \phantom{-}3.0383786892 \\
H$_7$ & -0.1111367729 & \phantom{-}1.7667671082 & \phantom{-}3.5399272616 \\
H\ud & \phantom{-}0.1977894112 & -0.3640813188 & \phantom{-}0.9804101222 \\
H\ua & \phantom{-}0.2837241404 & -0.2724755806 & \phantom{-}1.5402824816 \\
\end{tabular}
\begin{tablenotes}
\item[\textit{a}]\footnotesize All values are in \AA.
\end{tablenotes}
\end{threeparttable}

\vfill
\clearpage

\null
\vfill

\centering
\begin{threeparttable}
\captionof{table}{\justifying NEO-MECP for D transfer in CBO\tnote{\textit{a}}}
\begin{tabular}{c c c c}\label{tab:si_cbo_deut_mcep}
Atom & X & Y & Z \\
\hline
O$_1$ & \phantom{-}0.3735270695 & -0.0020014111 & \phantom{-}2.5748141844 \\
C$_1$ & -0.1119488102 & \phantom{-}0.0019360621 & -0.1013056433 \\
C$_2$ & -0.4625774530 & \phantom{-}1.0771296257 & \phantom{-}2.7298214251 \\
C$_3$ & -0.6242798184 & \phantom{-}1.9339718386 & \phantom{-}1.4449684024 \\
C$_4$ & -1.1321408181 & \phantom{-}1.1103774083 & \phantom{-}0.2524432558 \\
C$_5$ & -0.6213972932 & -1.0222909676 & -0.9803223672 \\
N$_1$ & -1.0370712886 & -1.8738716157 & -1.6673828474 \\
H$_1$ & \phantom{-}0.3541139194 & \phantom{-}2.3799100815 & \phantom{-}1.1912011411 \\
H$_2$ & -1.3241579200 & \phantom{-}2.7666631468 & \phantom{-}1.6469512836 \\
H$_3$ & -1.3496205405 & \phantom{-}1.7607560545 & -0.6146732675 \\
H$_4$ & -2.0859692917 & \phantom{-}0.6345173145 & \phantom{-}0.5346571193 \\
H$_5$ & \phantom{-}0.7975503294 & \phantom{-}0.4351827107 & -0.5463878592 \\
H$_6$ & -1.5073768791 & \phantom{-}0.7810317798 & \phantom{-}3.0367364557 \\
H$_7$ & -0.1174755288 & \phantom{-}1.7718961471 & \phantom{-}3.5421728894 \\
D\ud & \phantom{-}0.1953197104 & -0.3741951051 & \phantom{-}0.9605713787 \\
D\ua & \phantom{-}0.2907100680 & -0.2719025172 & \phantom{-}1.5617111201 \\
\end{tabular}
\begin{tablenotes}
\item[\textit{a}]\footnotesize All values are in \AA.
\end{tablenotes}
\end{threeparttable}

\vfill
\clearpage

\clearpage
\phantomsection
\addcontentsline{toc}{section}{References}
\providecommand{\latin}[1]{#1}
\makeatletter
\providecommand{\doi}
  {\begingroup\let\do\@makeother\dospecials
  \catcode`\{=1 \catcode`\}=2 \doi@aux}
\providecommand{\doi@aux}[1]{\endgroup\texttt{#1}}
\makeatother
\providecommand*\mcitethebibliography{\thebibliography}
\csname @ifundefined\endcsname{endmcitethebibliography}
  {\let\endmcitethebibliography\endthebibliography}{}

\end{document}